\documentclass[preprint,12pt,authoryear,table]{elsarticle}

\date{}
\usepackage{amsthm,amssymb,amsfonts}
\usepackage{amsmath}
 \usepackage{graphicx}
\usepackage{booktabs}
\usepackage{float}
\usepackage{comment}
\usepackage{natbib}
\usepackage{xfrac}
\usepackage{adjustbox}
\usepackage{lscape}
\usepackage{listings}
\usepackage[ruled,vlined]{algorithm2e}
\usepackage{mathrsfs} 
\usepackage{tikz} 
\usepackage[skip=0.5\baselineskip]{caption}

\usepackage{marginnote,datetime,enumitem,subfigure,rotating,fancyvrb}
\usepackage{longtable, multirow, colortbl, graphicx, array}
\usepackage{xcolor}
\usepackage[margin=1.3in]{geometry} 

\usepackage{algorithmic}
\usepackage{wrapfig}
\input xy
\xyoption{all}

\usepackage[colorlinks,pagebackref=true]{hyperref}
\begin{document}

\begin{frontmatter}


\title{Optimizing Portfolio with Two-Sided Transactions and Lending: A Reinforcement Learning Framework}

\author[vt]{Ali Habibnia}
\ead{habibnia@vt.edu}
\author[epfl]{Mahdi Soltanzadeh}
\ead{mahdi.sltz@gmail.com}

\address[vt]{Department of Economics, Virginia Tech}
\address[epfl]{Department of Economics and Finance, Tehran Institute for Advanced Studies (TeIAS)}

\tnotetext[t1]{}

\begin{abstract}
This study presents a Reinforcement Learning (RL)-based portfolio management model tailored for high-risk environments, addressing the limitations of traditional RL models and exploiting market opportunities through two-sided transactions and lending. Our approach integrates a new environmental formulation with a Profit and Loss (PnL)-based reward function, enhancing the RL agent's ability in downside risk management and capital optimization. We implemented the model using the Soft Actor-Critic (SAC) agent with a Convolutional Neural Network with Multi-Head Attention (CNN-MHA). This setup effectively manages a diversified 12-crypto asset portfolio in the Binance perpetual futures market, leveraging USDT for both granting and receiving loans and rebalancing every 4 hours, utilizing market data from the preceding 48 hours. Tested over two 16-month periods of varying market volatility, the model significantly outperformed benchmarks, particularly in high-volatility scenarios, achieving higher return-to-risk ratios and demonstrating robust profitability. These results confirm the model's effectiveness in leveraging market dynamics and managing risks in volatile environments like the cryptocurrency market.
\end{abstract}

\end{frontmatter}



\newpage

\section{Introduction}\label{sec:intro}

The escalating complexity of today's financial markets, characterized by the globalization of financial activities, the rapid growth in the number and types of financial products (such as derivatives and complex investment vehicles), the increased use of technology and algorithmic trading, and the interconnectedness of global economies and financial systems, has challenged the adequacy of traditional portfolio management models. They struggle with adapting to new market features and complexities, relying on single-period or oversimplified multi-period modeling. This shortfall is particularly noticeable in highly non-stationary markets. \cite{li2023online}.

To address these challenges, machine learning (ML) approaches have been increasingly applied to portfolio optimization e.g. \cite{mulvey2020}, \cite{barua2022ml}, \cite{lin2023ml}, \cite{liu2023ml} and \cite{barua2023ml}. However, these ML models often depend on predicting price trends, which, as shown in various studies like \cite{moody2001direct}, may not necessarily correlate with maximizing portfolio performance. This realization has sparked interest in exploring model-free approaches, such as Reinforcement Learning (RL), known for their efficacy in policy-based problems across diverse domains \cite{cui2023economicm}. RL-based portfolio management models uniquely integrate prediction and weight management into a unified process and are capable of optimizing portfolio performance for varied objectives and constraints.

Expanding on the definition of RL by \cite{sutton2018reinforcement}, a key aspect of portfolio management within this framework lies in the understanding of the environment states, which is analogous to the historical data of assets. Decisions or actions in this context involve transactions to rebalance the assets' weights in the portfolio. The crux of these actions is underpinned by the reward, defined as scalar functions of the financial earnings or losses resulting from these transactions.

There have been significant advances in applying RL to the portfolio management problem in recent years, but the field has not fully explored the effective use of new market facilities in the model. Integrating advanced market features like two-sided transactions and lending into RL adds considerable risk and complexity, resulting in the response's non-convergence or undesirable outcomes. Our study aims to tackle this issue by developing a new environmental formulation with a Profit and Loss (PnL)-based reward function. By focusing on key constraints like transaction costs, and employing the Soft Actor-Critic (SAC) agent with a Convolutional Neural Network with Multi-Head Attention (CNN-MHA) architecture, we enhance the model's ability to downside risk management and capital optimization.

The SAC agent (\cite{haarnoja2018soft}), a model-free, actor-critic algorithm, aims to maximize both reward and entropy, with entropy serving as a gauge for policy uncertainty. This dual focus fosters an equilibrium between reward maximization and environmental exploration, granting the agent the resilience to effectively adapt to novel market scenarios. 
The MHA architecture (\cite{vaswani2017attention}) complements the SAC agent by transcending the constraints of traditional Recurrent Neural Networks (RNNs). The MHA architecture relies on the self-attention mechanism, which enables comprehensive contextual processing of sequential data \cite{li2023attention}. This mechanism allows each element in a sequence to relate to every other element, effectively overcoming the limitations of dependency in previous sequence networks like Long Short-Term Memory (LSTM).

In the implementation phase, we utilize a diversified 12-crypto asset portfolio in the Binance perpetual futures market as the RL environment to train and test our model. The cryptocurrency market, with unique characteristics of continuous 24-hour transactions, decentralization, low transaction fees, high security through blockchain technology \cite{fang2022}, and significant tail-risk exposure \cite{borri2019}, presents a challenging but appropriate landscape for implementing and evaluating ML-based portfolio management models \cite{ren2022past}.

The remainder of this paper is organized as follows: Section \ref{sec:literature} explores existing portfolio management models in literature. Section \ref{Sec:methodology} details the research methods and data used. Section \ref{sec:emp} presents the empirical results, including an analysis of test results compared to benchmark models. The paper concludes with a summary of the study's findings in Section \ref{Sec:conclusion}.

\section{Literature Review} \label{sec:literature}

In recent years, building on \cite{marko1952}'s research, numerous studies have focused on mathematical modeling to tackle the portfolio management problem. These efforts can be divided into two main categories: Single-Period Portfolio Optimization (SPPO) and Multi-Period Portfolio Optimization (MPPO) models.

SPPO models adopt a static approach, optimizing portfolio weights at the outset of the investment period without further modifications. A significant evolution in this category is the introduction of alternative risk measures in place of variance, as proposed in the original Markowitz framework. Innovations include Variance with Skewness (VwS) by \cite{samuelson1958}, Semi-Variance (S-V) by \cite{markovitz1959}, Mean-Absolute Deviation (MAD) by \cite{konno1991}, Value-at-Risk (VaR) by \cite{jorion1997}, Minimax (MM) by \cite{young1998}, and Conditional Value-at-Risk (CVaR) by \cite{rockafellar2000}. These models typically employ quadratic or linear programming for Efficient Frontier (EF) determination, with several studies have explored multi-objective optimization methods to discover a more optimal EF, e.g., \cite{roman2007multi-obj}, \cite{tsao2010multi-obj}, \cite{petchrompo2022multi-obj} and \cite{wang2023multi-obj}.
However, SPPO models are often critiqued for their reliance on historical data and oversimplified assumptions (\cite{michaud1989}), limiting their applicability in multi-stage and long-term investment scenarios.

Conversely, MPPO models, adhering to \cite{kelly1956}’s optimal capital growth approach, allow periodic reassessment and readjustment of portfolio weights. This category, which began with the contributions of  \cite{smith1967}, \cite{mossin1968}, \cite{merton1969}, \cite{chen1971} and, \cite{fama1975} researches and was further advanced by \cite{li2000} and \cite{zhou2000}, explored in both continuous and discrete time frameworks, leverages Dynamic Programming (DP), Stochastic Programming (SP), and Stochastic Dynamic Programming (SDP) to find optimal portfolio weights.

DP-based MPPO models, ideal for long-term investments with frequent rebalancing, face computational complexity and reliance on simplifications due to practical constraints like transaction costs, e.g., \cite{kraft2013dynamic}, \cite{cui2017} and \cite{das2022dynamic}. On the other hand, SP-based models, incorporating practical constraints, offer a more realistic approach but face computational challenges in long-term scenarios, e.g., \cite{gaivoronski2003sp}, \cite{fulton2019sp} and \cite{barro2022}. SDP-based models, including RL, integrate the benefits of DP and SP approaches, making them suitable for long-term investment strategies involving random variables, e.g., \cite{topaloglou2008}, \cite{luo2014sdp}, and \cite{van2018}.

The application of RL in portfolio management models first gained attention through \cite{neuneier1997}, which utilized the Q-Learning agent to identify optimal portfolio management policy. Following this, \cite{Mihatsch1998} enhanced the Q-Learning by incorporating risk assessment. Later on \cite{gao2000} developed a Q-Learning-based portfolio management model aimed at maximizing the Sharpe ratio. Further advancements were made by \cite{moody2001direct}, who introduced Recurrent Reinforcement Learning (RRL) for portfolio management, utilizing the Direct Reinforcement (DR) algorithm.

Recent contributions in this area can be divided into the five key elements of an RL-based model: the Agent, the Environment, Actions, States, and the Reward function, focused on improving weight optimization.

\textit{The agent}, comprises two essential components: the policy and the learning algorithm. The policy, often implemented through Neural Networks (NN), determines portfolio weights based on the current state. Notable recent research in NN design includes articles \cite{li2023online}, and \cite{zhao2023corr}, with a focus on the use of attention mechanism. The learning algorithm updates the policy using a trajectory of states, actions, and rewards. Popular learning algorithms in portfolio management, as outlined in \cite{hambly2023recent}, include Q-learning, DQN, DDQN, DPG, DDPG, and A2C, with several articles like \cite{halperin2019}, \cite{jaisson2022}, \cite{aboussalah2022}, \cite{coache2023}, and \cite{mounjid2023} offering customized learning algorithms for financial contexts.

\textit{The environment} integrates assets and market characteristics with a defined strategy for portfolio optimization. Notable recent developments in this field include \cite{wang2020continuous}'s implementation of the mean-variance model in a continuous-time format using RL, \cite{wu2021emn}'s formulation of an environment based on the Equity Market Neutral (EMN) strategy, \cite{jin2023meanVar}'s implementation of the mean-VaR strategy with two-sided transactions using RL, \cite{liu2023nonlinear}'s employment of RL to discover nonlinear relationships and refine pairs trading strategies, and \cite{cui2023economicm}'s efforts to reduce the tail risk of digital currency portfolios by integrating the mean-CVaR strategy with RL.

\textit{Actions} or optimal portfolio weights are divided into discrete (buy, hold, sell) and continuous (direct weight determination) approaches, based on the characteristics of the agent. Discrete actions, though limited in precision, often incorporate complex reward functions with multiple constraints to address this limitation, as utilized in \cite{jaisson2022}, and \cite{ngo2023does}. Conversely, continuous actions improve accuracy by allowing the model to directly optimize portfolio weights. However, the majority of research using this approach does not account for negative weights and tends to concentrate on the spot market, like \cite{betancourt2021}, and \cite{li2023online} works.

\textit{states} often refer to historical asset data. In the financial ML context, data preprocessing aimed at enhancing model performance involves signal decomposition techniques like Discrete Wavelet Transform (DWT) and Empirical Mode Decomposition (EMD), as shown in articles \cite{cheng2014novel_decom}, \cite{wang2018crude_decom}, \cite{faria2018forecasting_decom}, \cite{lee2021wavelet}, \cite{xu2023limited_decom}, \cite{tangstock_decom}, \cite{stein2024forecasting_decom} and the application of technical indicators, similar to \cite{gradojevic2013fuzzy_tec}, \cite{liu2022stock_tec}, \cite{ma2022deep_tec}, \cite{jang2023ti}, \cite{tan2023trend_tec} and \cite{goutte2023deep_tec}, in addition to data normalization.

The \textit{reward function} serves as the objective in optimizing portfolio weights. Portfolio return, Sharpe ratio, and PnL, as noted in \cite{hambly2023recent}, are commonly used reward functions in the literature. Additionally, several studies, such as \cite{halperin2019}, \cite{jaisson2022}, and \cite{aboussalah2022}, have developed specialized reward functions to optimize specific goals.

The literature review reveals a gap in research on negative weights and lending in RL-based portfolio management models, attributed to the increased risk these features introduce. This gap suggests a need for redesigning the environment and reward functions to enable the model to manage risk and capital more effectively. However, most existing studies opt for standard environments and conventional reward functions, focusing on optimizing networks and learning algorithms to improve performance.

\pagebreak
\section{Methodology}\label{Sec:methodology}

\subsection{Reinforcement Learning} \label{Subsec:RL}
RL can be conceptualized as a tuple $(S,\ A,\ R_a,\ P_a,\ \gamma)$, comprising:
\begin{itemize}
  \item $S$: The set of state representations.
  \item $A$: The set of actions available to the agent.
  \item $R_a$: The immediate reward received after transitioning to a new state due to action $a$.
  \item $P_a$: The state transition probability, expressed as $P_a(s^\prime,s)=P_r(s_{t+1}=s^\prime \mid s_t=s,a_t=a)$, indicating the likelihood of transitioning to state $s^\prime$ from state $s$ due to action $a$ at time $t$.
  \item $\gamma$: The discount factor (0 to 1) signifies the importance of past actions.
\end{itemize}

\begin{figure}[H]
    \centering
    \includegraphics[scale=.9]{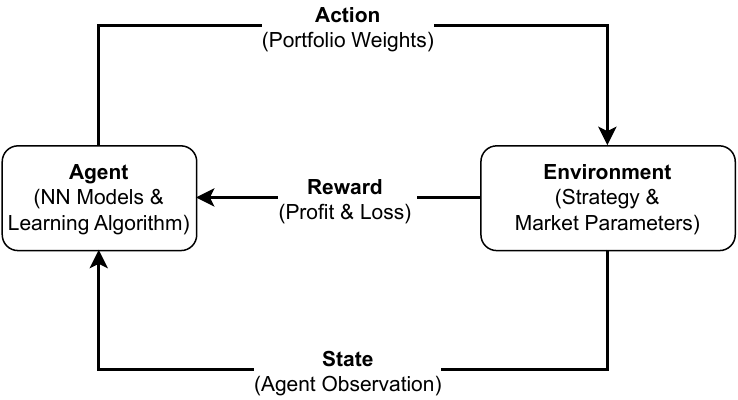}
    \caption{Architecture of reinforcement learning in portfolio management.}
    \label{fig:rl}
\end{figure}

The agent aims to find an optimal policy mapping states to actions to maximize cumulative discounted rewards, as defined in \cite{cui2023economicm}. Figure \ref{fig:rl} illustrates the components of the RL-based portfolio management model.


\subsection{Data Preprocessing} \label{Subsec:data}
Given the non-stationarity and varied scales of asset prices, preprocessing is necessary before network application. This process (Equation \ref{eq:1}) involves normalizing and converting price data into a tensor ($T$), with $M\times4\times N$ dimension in each state. Here, $M$ represents the number of portfolio assets, 4 the number of features (open, high, low, and close prices), and $N$ the historical data length for each feature.

\begin{equation} \label{eq:1}
\begin{aligned}
& T_t=\left[P_t^1,\ {\ P}_t^2,\ \ldots,\ {\ P}_t^{M-1},\ {\ P}_t^M\right] \\
& P_t^i=\left[{Open}_t^i,\ {\ High}_t^i,\  {\ Low}_t^i,\ {\ Close}_t^i\right] \\
& {Open}_t^i=\left[0,\ \frac{{open}_{t-n+2}^i}{{open}_{t-n+1}^i}-1,\ \ldots,\ \frac{{open}_{t-1}^i}{{open}_{t-n+1}^i}-1,\frac{{open}_t^i}{{open}_{t-n+1}^i}-1\right]+\ noise \\
& {High}_t^i=\left[0,\ \frac{{high}_{t-n+2}^i}{{high}_{t-n+1}^i}-1,\ \ldots,\ \frac{{high}_{t-1}^i}{{high}_{t-n+1}^i}-1,\frac{{high}_t^i}{{high}_{t-n+1}^i}-1\right]+\ noise \\
& {Low}_t^i=\left[0,\ \frac{{low}_{t-n+2}^i}{{low}_{t-n+1}^i}-1,\ \ldots,\ \frac{{low}_{t-1}^i}{{low}_{t-n+1}^i}-1,\frac{{low}_t^i}{{low}_{t-n+1}^i}-1\right]+\ noise \\
& {Close}_t^i=\left[0,\ \frac{{close}_{t-n+2}^i}{{close}_{t-n+1}^i}-1,\ \ldots,\ \frac{{close}_{t-1}^i}{{close}_{t-n+1}^i}-1,\frac{{close}_t^i}{{close}_{t-n+1}^i}-1\right]+\ noise
\end{aligned}
\end{equation}

$T_t$ comprises $M$ price vectors ($P_t$) for each asset at period $t$, with $P_t^i$ representing the price vector for asset $i$ at period $t$, including four dimensions, ${Open}_t^i$, ${High}_t^i$, ${Low}_t^i$, and ${Close}_t^i$. These dimensions are normalized to reflect relative price changes, with additional scaled normal noise introduced to make the agent more resilient to market noise.

\subsection{RL Agent} \label{Subsec: agent}
We employ the SAC agent, which utilizes two networks called the actor and the critic. the actor, determining the optimal weight rebalancing policy, while the critic guides the actor by evaluating environment rewards to maximize long-term discounted rewards and entropy. The actor's final layers output the mean and standard deviation of the action's normal distribution, from which the optimal weights are randomly selected. However, for model testing, we select actions with the highest likelihood as the portfolio weights.

\begin{figure}[H]
    \centering
    \includegraphics[scale=.69]{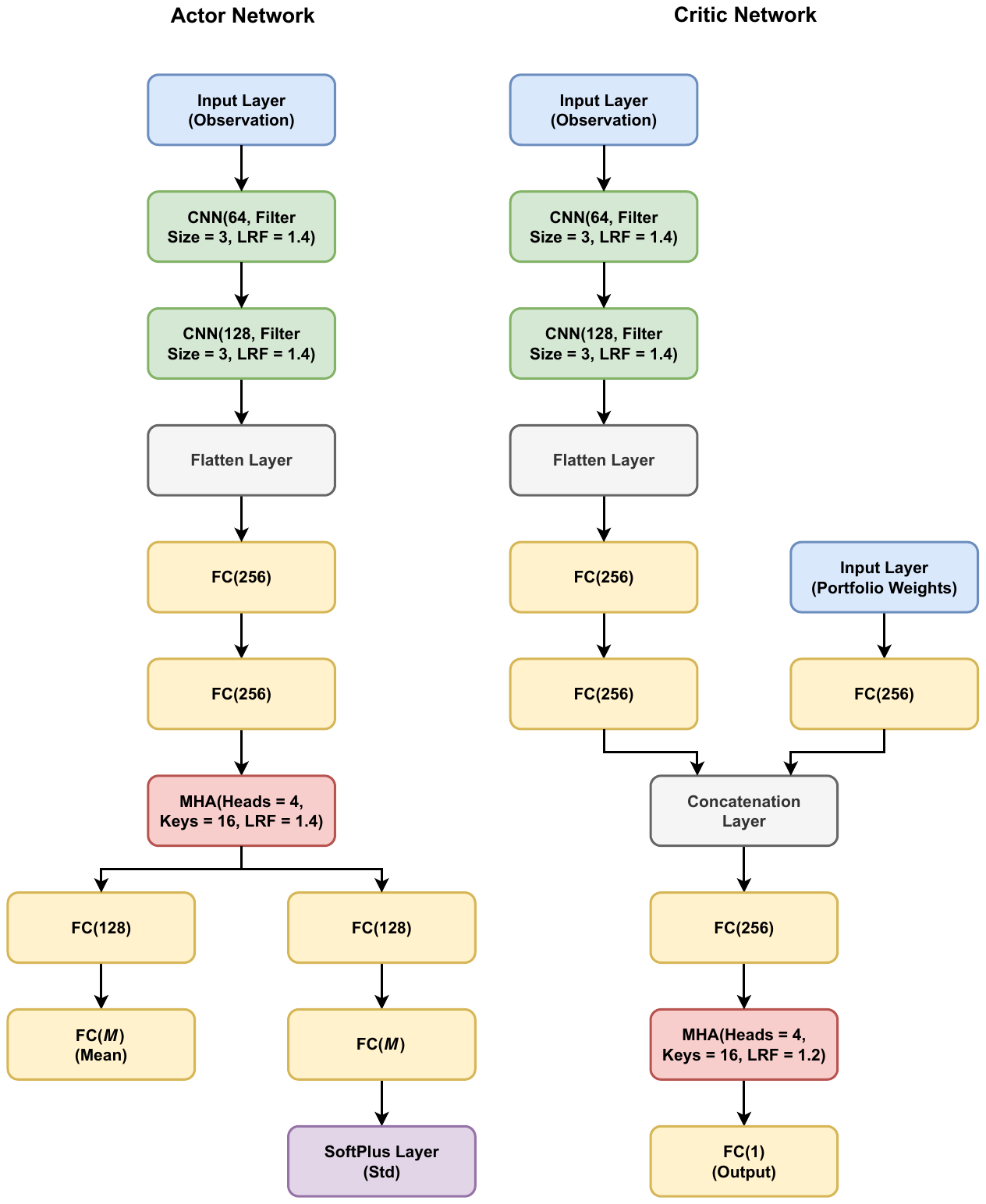}
    \caption{The architecture of the actor and critic networks.}
    \label{fig:actorcritic}
\end{figure}

In the development of the deep NN architectures for an $M$ asset portfolio, as illustrated in Figure \ref{fig:actorcritic}, we enhanced the input data by incorporating two CNN layers equipped with 64 and 128 filters, respectively, for feature extraction. Additionally, integrating an MHA layer, which highlights the more important parts of the sequence through an attention mechanism, boosts the agent's efficiency. In addition, the designed architectures utilize Fully Connected (FC) layers, incorporate the LeakyReLU activation function across all intermediate layers, and apply a unique optimized Learning Rate Factor (LRF) for each layer.


\subsection{RL Environment} \label{Subsec:environment}
The RL environment primarily undertakes two key tasks: calculating the reward associated with each action taken and transitioning the agent to the next state through the defined strategy.

\pagebreak


At each state, the agent rebalances the portfolio weights using a vector $W_t$ with $1\times\left(M+1\right)$ dimension. In this vector, $\forall w_t^i\in W_t,w_t^i\in\left[-1,1\right]$, $\sum_{i=1}^{m}\left|w_t^i\right|=1$, and $w_t^{m+1}$ signifies the loan amount. 
A positive value of $w_t^{m+1}$ indicates the allocation of a portion of cash as a loan to accrue interest, while a negative value implies taking out a loan, increasing cash in the portfolio, and incurring interest.

When opening or closing positions in exchanges and brokers, a transaction fee ($\delta$) is deducted. However, typically, there are no fees associated with either granting or receiving loans, and only the interest rate ($\zeta$) applies. With these explanations, portfolio rebalancing calculations are presented by Equations \ref{eq:2} to \ref{eq:7}.

For each state, the available rebalancing capital ($c_t$) is determined by multiplying $w_t^{m+1}$  by the portfolio value at time $t$ ($p_t$).

\begin{equation} \label{eq:2}
c_t=p_t\times\left(1-w_t^{m+1}\right)
\end{equation}

Next, the interest incurred from either granting (positive value) or receiving (negative value) a loan over the rebalancing interval is computed.

\begin{equation} \label{eq:3}
{interest}_t=\ p_t\times w_t^{m+1}\times\zeta
\end{equation}


Given the current asset values in the portfolio (${Port}_{t-1}$), the necessary changes according to $W_t$, and the related transaction costs (${Cost}_t$), the readjusted asset values (${Port}_t^\prime$) are calculated.

\begin{equation} \label{eq:4}
{Port}_t^\prime=c_t\times W_t^{1:m}-{Cost}_t
\end{equation}

\begin{equation} \label{eq:5}
{Cost}_t=\delta\times\left(c_t\times W_t^{1:m}-{Port}_{t-1}\right)
\end{equation}

Finally, the ${Port}_t$ value in the next state is calculated by multiplying ${Port}_t^\prime$ by the assets’ return ($R_t$),

\begin{equation} \label{eq:6}
{Port}_t={Port}_t^\prime
\times\left(1+R_t\right)
\end{equation}

\noindent and the portfolio value at time $t+1$ is obtained as $p_{t+1}$.
 
\begin{equation} \label{eq:7}
p_{t+1}=p_t+\ \sum{\left({Port}_t^\prime\times R_t\right)}+{interest}_t
\end{equation}

The environment also computes the reward associated with the agent's action to transition to the next state. The reward function, as outlined in Equation \ref{eq:8}, is designed to update the agent's policy to maximize rebalancing profit while controlling losses during training. Furthermore, to prevent the agent from making excessive modifications to the portfolio weights, transaction costs are incorporated into the reward function.

\begin{equation} \label{eq:8}
{reward}_t=\frac{\sum{Profit}_t-penalty\times\left(\sum
{Loss}_t+\sfrac{1}{5}\times\sum{Cost}_t\right)}{\ c_t}
\end{equation}

The $penalty$ parameter in the reward function is also intended to modulate the significance of losses and transaction costs. It can be set according to the investor's risk aversion level.

\subsection{General Assumptions} \label{Subsec:assumptions}

In many research, including \cite{kwan1999note}, \cite{zhang2020} and \cite{zhao2023corr}, two core assumptions are typically made for such models:

\begin{itemize}
\item \textbf{Full Liquidity:} Transactions are assumed to be executed instantly upon decision.
\item \textbf{Market Neutrality:} Transactions are assumed not to affect future asset prices.
\end{itemize}

Furthermore, although collateral is usually required for loans, we assume it is unnecessary in our context.

\subsection{Benchmark Models} \label{Subsec:benchmark}


We employ return as the reward function to benchmark our model using the same agent and environment. also, the use of return-to-risk based reward functions is precluded due to lending opportunities in the environment.

Additionally, we select three benchmarks from traditional SPPO models, inspired by \cite{ngo2023does} and \cite{jin2023meanVar}. SPPO models can be compared to our model by applying weight rebalancing in each state of the environment. Differences in risk calculation lead to distinct EFs, with optimal weights determined at the intersection of the EF and the investor's utility function or $w_t^{m+1}=1$ if no intersection exists.

\subsubsection{Return-based RL} \label{Subsubsec:ret-based-rl}
Return (Equation \ref{eq:return}) is a favored reward function in the RL-based portfolio management models due to its simplicity, adaptability to various environments, and alignment with the primary goal of maximizing the portfolio value ($p$).

\begin{equation} \label{eq:return}
Return\ =\ \frac{p_t}{p_{t-1}}-1
\end{equation}


\subsubsection{Mean-Variance (MV)} \label{Subsubsec:mv}

The MV model (\cite{marko1952}) is a computational framework used to select portfolio weights based on assessing risk to the expected return. In this model, portfolio risk (variance) is calculated using Equation \ref{eq:9}.

\begin{equation} \label{eq:9}
\sigma_P=\sum_{i} w_i^2\sigma_i^2+\sum_{i}\sum_{j\neq i} w_iw_j\sigma_i\sigma_j\rho_{ij}
\end{equation}

where $w_i$ and $\sigma_i$ are the weight and standard deviation of the $i$th asset, respectively, and $\rho_{ij}$ is the correlation coefficient between assets $i$ and $j$. Additionally, the EF in this model is derived using quadratic programming.

\subsubsection{Mean Absolute Deviation (MAD)} \label{Subsubsec:mad}
MAD, a risk measure introduced by \cite{konno1991} for portfolio management, quantifies the mean absolute deviation of individual asset returns from the portfolio's average return. It captures return dispersion and provides a robust risk measure less affected by extreme values. MAD is calculated using Equation \ref{eq:10}, and the EF is determined using linear programming.

\begin{equation} \label{eq:10}
{MAD}_P=\frac{1}{n}\sum_{i=1}^{n}\left|r_i-E\left(r_P\right)\right|
\end{equation}

where $n$ is the number of assets in the portfolio, $r_i$ and $E\left(r_P\right)$ are the return of the $i$th asset and the mean return of the portfolio, respectively.

\subsubsection{Conditional Value at Risk (CVaR)} \label{Subsubsec:cvar}
CVaR is a risk measure that assesses potential losses beyond a specific confidence level. It quantifies the expected loss exceeding the confidence level and highlights extreme loss severity, focusing on tail risk. CVaR aids in making informed decisions to mitigate potential downside risks in the portfolio management model introduced by \cite{rockafellar2000}. This model employs quadratic programming to derive the EF, with CVaR being determined through Equations \ref{eq:11} and \ref{eq:12}.

\begin{equation} \label{eq:11}
CVaR=E\left(r_P\middle| r_P\geq{VaR}^\alpha\right)
\end{equation}

\begin{equation} \label{eq:12}
{VaR}^\alpha=F^{-1}\left(1-\alpha\right)
\end{equation}

where ${VaR}^\alpha$ represents the maximum expected loss with confidence level $\alpha$, assuming portfolio returns ($r_P$) follow a normal distribution.

\section{Empirical Studies}\label{sec:emp}

\subsection{Setup Environment and Agent} \label{Subsec:setup}

In the model's implementation phase, we chose a diversified 12-crypto asset portfolio from the Binance perpetual futures market based on their liquidity and data availability. USDT serves as the portfolio's base asset, capable of granting and receiving loans. Our analysis covers two distinct periods: a high-volatility period from May 1, 2021, to September 1, 2022 (Port A), and a low-volatility period from June 1, 2022, to October 1, 2023 (Port B). Each period lasts 16 months, divided into 12 months (2178 steps) for training and 4 months (732 steps) for testing, featuring 4-hour rebalancing intervals. Additionally, hourly historical price data for cryptocurrencies is sourced from \url{www.cryptodatadownload.com}.

\begin{figure}[H]
    \centering
    \includegraphics[scale=.75]{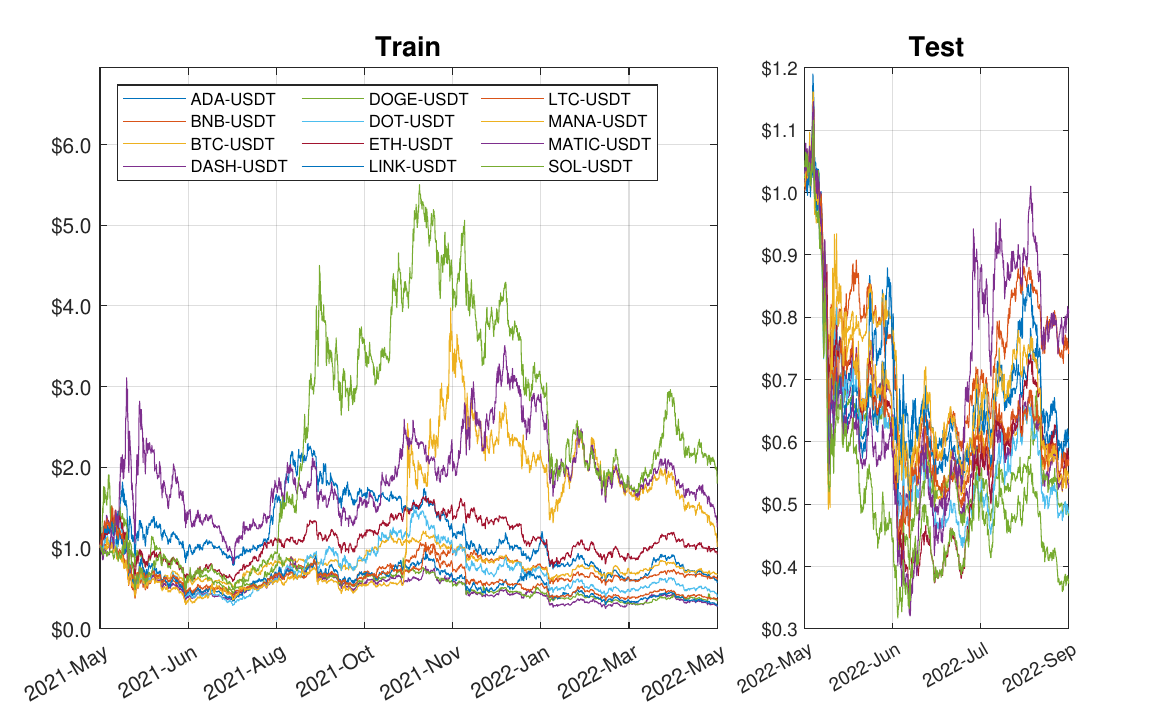}
    \caption{Investment value of one dollar across each currency in Port A during training and testing.}
    \label{fig:3}
\end{figure}

\begin{figure}[H]
    \centering
    \includegraphics[scale=.75]{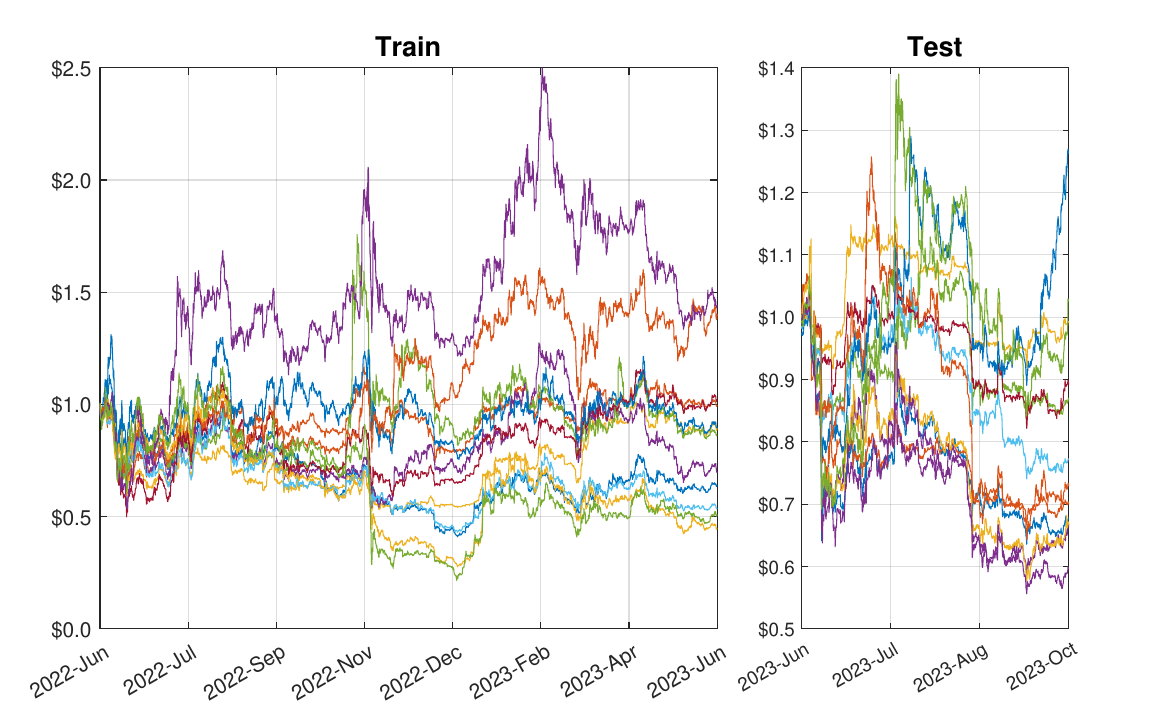}
    \caption{Investment value of one dollar across each currency in Port B during training and testing.}
    \label{fig:4}
\end{figure}

Figures \ref{fig:3} and \ref{fig:4} illustrate the investment value of one dollar across each currency within Portfolios A and B, respectively, during training and testing, and Table \ref{tab:1} provides statistical data on the 4-hour return distributions of the currencies.

\begin{landscape}
\begin{table}[H]
\centering
\caption{Statistics for 4-hour return distributions of the currencies in Portfolios A and B.}
\label{tab:1}
\small
\begin{tabular}{ccccccccccc}
\hline
\multirow{3}{*}{\textbf{Currencies INFO}} & Mean(\%) & Std(\%) & Min(\%) & Max(\%) & Total(\%) & Mean(\%) & Std(\%) & Min(\%) & Max(\%) & Total(\%) \\ \cline{2-11} 
                                 & \multicolumn{10}{c}{\textit{Port A}}                                                                               \\ \cline{2-11} 
                                 & \multicolumn{5}{c}{Train}                           & \multicolumn{5}{c}{Test}                            \\ \hline
ADA-USDT                         & 0.00     & 2.37    & -11.09  & 12.70   & -43.35     & -0.04    & 2.37    & -14.88  & 15.45   & -41.77     \\
BNB-USDT                         & 0.00     & 2.09    & -13.82  & 10.59   & -39.40     & -0.02    & 1.82    & -6.19   & 14.34   & -27.18     \\
BTC-USDT                         & -0.01    & 1.59    & -10.40  & 8.22    & -34.75     & -0.07    & 1.54    & -6.61   & 7.59    & -47.12     \\
DASH-USDT                        & -0.03    & 2.56    & -13.40  & 16.46   & -73.30     & -0.06    & 2.16    & -10.47  & 11.68   & -48.77     \\
DOGE-USDT                        & -0.01    & 2.98    & -15.77  & 39.31   & -66.51     & -0.08    & 2.09    & -11.56  & 12.85   & -52.54     \\
DOT-USDT                         & 0.00     & 2.79    & -24.95  & 21.46   & -61.36     & -0.07    & 2.37    & -10.83  & 13.06   & -52.64     \\
ETH-USDT                         & 0.02     & 2.05    & -10.68  & 9.67    & -10.03     & -0.05    & 2.15    & -7.57   & 10.42   & -44.04     \\
LINK-USDT                        & -0.02    & 2.74    & -13.69  & 13.26   & -72.45     & -0.04    & 2.41    & -9.10   & 9.38    & -40.27     \\
LTC-USDT                         & -0.02    & 2.28    & -14.78  & 12.27   & -64.87     & -0.06    & 2.07    & -8.49   & 10.43   & -44.01     \\
MANA-USDT                        & 0.06     & 3.45    & -17.03  & 41.11   & 0.94       & -0.04    & 3.09    & -10.85  & 42.23   & -46.33     \\
MATIC-USDT                       & 0.06     & 3.21    & -15.37  & 21.31   & 25.92      & 0.01     & 2.87    & -12.34  & 13.82   & -20.92     \\
SOL-USDT                         & 0.07     & 3.01    & -16.90  & 17.12   & 77.00      & -0.09    & 2.82    & -13.92  & 13.89   & -64.18     \\ \hline
                                 & \multicolumn{10}{c}{\textit{Port B}}                                                                               \\ \hline
ADA-USDT                         & -0.01    & 1.61    & -7.52   & 10.80   & -36.39     & -0.04    & 1.40    & -13.17  & 11.76   & -30.10     \\
BNB-USDT                         & 0.01     & 1.27    & -5.74   & 11.74   & 0.02       & -0.04    & 0.96    & -8.39   & 5.00    & -29.42     \\
BTC-USDT                         & 0.00     & 1.19    & -7.52   & 8.67    & -11.02     & 0.00     & 0.72    & -4.59   & 5.58    & 0.67       \\
DASH-USDT                        & 0.00     & 1.71    & -12.30  & 13.08   & -28.16     & -0.04    & 1.47    & -18.53  & 6.43    & -32.96     \\
DOGE-USDT                        & 0.01     & 2.06    & -12.73  & 29.36   & -12.99     & -0.01    & 1.26    & -11.33  & 8.30    & -13.06     \\
DOT-USDT                         & -0.02    & 1.60    & -8.72   & 10.27   & -45.87     & -0.03    & 1.09    & -9.91   & 5.68    & -21.35     \\
ETH-USDT                         & 0.01     & 1.59    & -10.19  & 11.35   & 3.00       & -0.01    & 0.75    & -5.21   & 4.54    & -9.76      \\
LINK-USDT                        & 0.01     & 1.82    & -13.54  & 11.89   & -7.60      & 0.04     & 1.41    & -11.58  & 12.14   & 28.22      \\
LTC-USDT                         & 0.03     & 1.75    & -14.58  & 15.76   & 41.39      & -0.03    & 1.43    & -13.04  & 13.19   & -27.63     \\
MANA-USDT                        & -0.02    & 1.95    & -10.47  & 21.59   & -54.23     & -0.04    & 1.46    & -11.02  & 7.36    & -33.17     \\
MATIC-USDT                       & 0.04     & 2.09    & -10.58  & 20.14   & 46.16      & -0.06    & 1.46    & -13.86  & 7.12    & -39.61     \\
SOL-USDT                         & 0.00     & 2.46    & -26.03  & 24.70   & -48.87     & 0.02     & 1.75    & -11.59  & 11.05   & 3.19       \\ \hline
\end{tabular}
\end{table}
\end{landscape}


Following the methodology in section \ref{Sec:methodology}, the environment's hyperparameters are detailed in Table \ref{tab:2}. We utilized real information from the Binance exchange and identified optimal parameter values for achieving satisfactory results. The environment is designed to rebalance portfolio weights every 4 hours, enhancing the profit-to-cost ratio. Shorter intervals resulted in increased trading costs without generating considerable additional profit, which is undesirable. Additionally, the analysis of the last 48 hours (i.e. $N=49$) of historical data enhanced the agent's market condition understanding. Utilizing data beyond 48 hours risked overemphasis on past market behavior. Furthermore, the penalty parameter in the reward function, aimed at controlling downside risk, is set to 25. This indicates the agent's sensitivity to loss is 25 times greater than profit.

\begin{table}[H]
\centering
\small
\caption{Environment's hyperparameters.}
\label{tab:2}
\begin{tabular}{lcc}
\cline{1-2}
Hyperparameters                               & Value    &  \\ \cline{1-2}
Initial investment                      & 1000\$   &  \\
Historical data length                  & 48 hours &  \\
Rebalancing frequency                   & 4 hours  &  \\
Penalty parameter                       & 25       &  \\
Annual interest rate for borrowing USDT & 5\%      &  \\
Annual interest rate for lending USDT   & 3\%      &  \\
Transaction fee                         & 0.05\%   &  \\
\cline{1-2}
\end{tabular}
\end{table}

The SAC agent's optimal hyperparameters are also outlined in Table \ref{tab:3}, fine-tuned to accelerate response convergence during training.

\begin{table}[H]
\centering
\small
\caption{The SAC agent's hyperparameters.}
\label{tab:3}
\begin{tabular}{lcc}
\cline{1-2}
Hyperparameters        & Value  &  \\ \cline{1-2}
Optimizer              & Adam   &  \\
Actor   Learning Rate  & 2e-4 &  \\
Critic   Learning Rate & 6e-4 &  \\
Discount   Factor      & 0.99   &  \\
Entropy   Weight       & 0.08   &  \\
Entropy   Learn Rate   & 6e-4 &  \\
L2RegularizationFactor & 1e-8  &  \\
Mini   Batch Size      & 16     &  \\
Sample   Time          & 1      &  \\
Steps   to Look Ahead  & 1      &  \\ \cline{1-2}
\end{tabular}
\end{table}

\subsection{Testing Results} \label{subsec:result}
\begin{figure}[H]
    \centering
    \includegraphics[scale=.7]{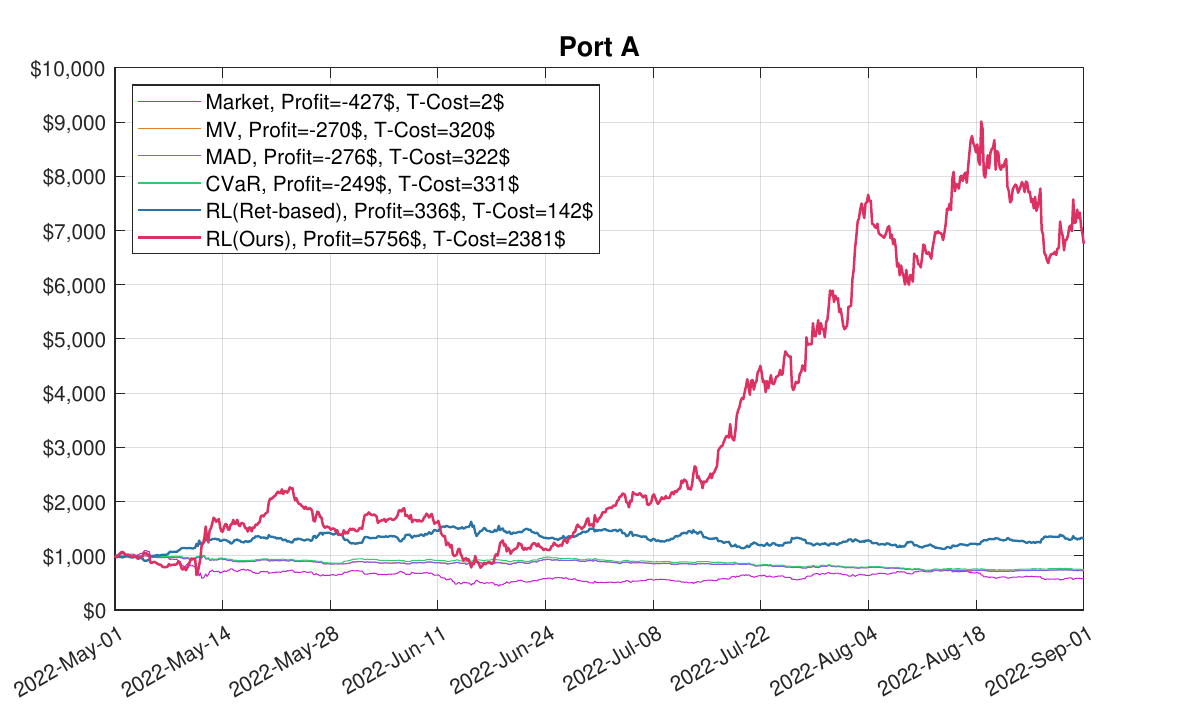}
    \caption{Testing results for Port A, Market represents an equally weighted strategy.}
    \label{fig:5}
\end{figure}

\begin{figure}[H]
    \centering
    \includegraphics[scale=.7]{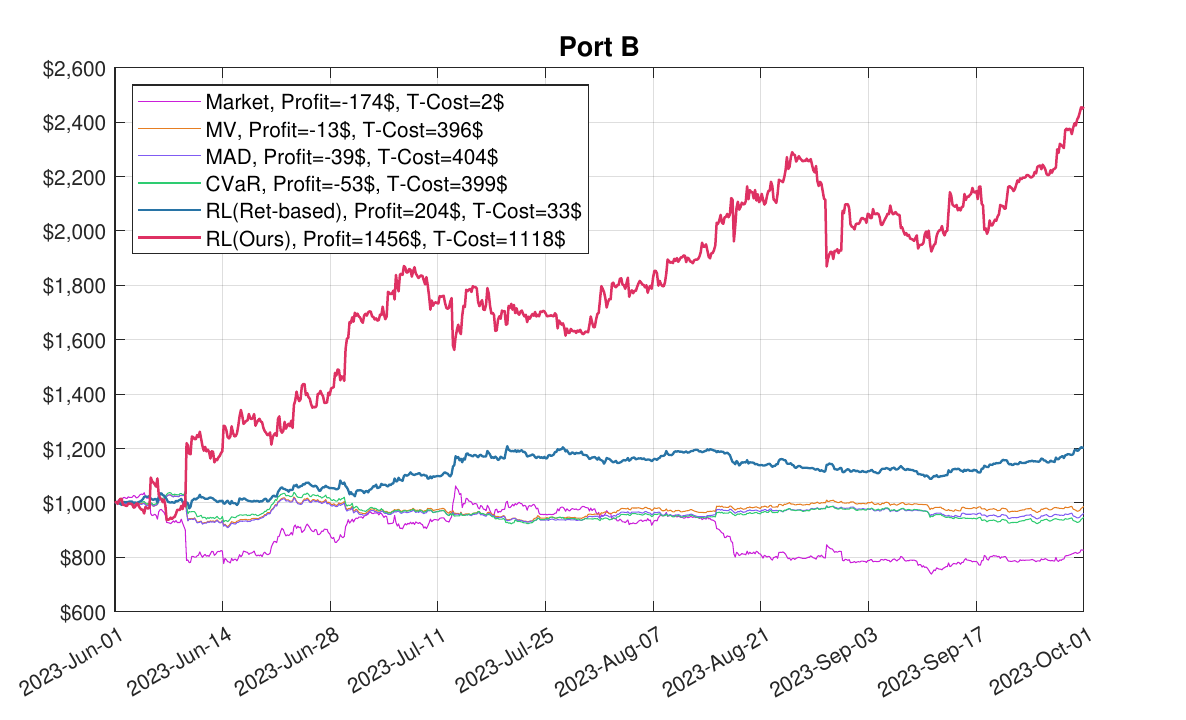}
    \caption{Testing results for Port B, Market represents an equally weighted strategy.}
    \label{fig:6}
\end{figure}

We tested our trained agent against return-based RL and SPPO benchmarks with a standard risk aversion parameter (i.e., $risk aversion=4$) as introduced in subsection \ref{Subsec:benchmark}. Figures \ref{fig:5} and \ref{fig:6} present the 4-month testing results for Portfolios A and B, respectively. Meanwhile, Table \ref{tab:4} provides a performance comparison based on the criteria outlined in the \cite{wood2021}'s paper.

\begin{table}[] 
\begin{center}
\caption{Comparison of the testing results for Portfolios A and B.}
\label{tab:4}
\small
\scalebox{0.67}{
\begin{tabular}{
>{\columncolor[HTML]{FFFFFF}}c 
>{\columncolor[HTML]{FFFFFF}}c 
>{\columncolor[HTML]{FFFFFF}}c 
>{\columncolor[HTML]{FFFFFF}}c 
>{\columncolor[HTML]{FFFFFF}}c 
>{\columncolor[HTML]{FFFFFF}}c 
>{\columncolor[HTML]{FFFFFF}}c 
>{\columncolor[HTML]{FFFFFF}}c 
>{\columncolor[HTML]{FFFFFF}}c 
>{\columncolor[HTML]{FFFFFF}}c }
\hline
\cellcolor[HTML]{FFFFFF}                                                                                        & \begin{tabular}[c]{@{}c@{}}Total\\  Return(\%)\end{tabular} & \begin{tabular}[c]{@{}c@{}}Win\\ Rate(\%)\end{tabular} & \begin{tabular}[c]{@{}c@{}}Average \\ Return(\%)\end{tabular} & \begin{tabular}[c]{@{}c@{}}Standard\\  Deviation(\%)\end{tabular} & \begin{tabular}[c]{@{}c@{}}Sharpe\\ Ratio\end{tabular} & \begin{tabular}[c]{@{}c@{}}Downside\\ Deviation(\%)\end{tabular} & \begin{tabular}[c]{@{}c@{}}Sortino\\  Ratio\end{tabular} & \begin{tabular}[c]{@{}c@{}}Max\\ Drawdown(\%)\end{tabular} & \begin{tabular}[c]{@{}c@{}}Calmar\\ Ratio\end{tabular} \\ \cline{2-10} 
\multirow{-2}{*}{\cellcolor[HTML]{FFFFFF}\textbf{\begin{tabular}[c]{@{}c@{}}Evaluation\\ Metrics\end{tabular}}} & \multicolumn{9}{c}{\cellcolor[HTML]{FFFFFF}\textit{Port A}}                                                                                                                                                                                                                                                                                                                                                                                                                                                                                                           \\ \hline
MV                                                                                                              & -26.978                                                     & 49.379                                                 & -0.040                                                        & \textbf{0.679}                                                    & -1.487                                                 & \textbf{0.513}                                                   & -1.970                                                   & 28.547                                                     & -0.961                                                 \\
{\color[HTML]{000000} MAD}                                                                                      & -27.604                                                     & 49.481                                                 & -0.042                                                        & 0.681                                                             & -1.518                                                 & 0.518                                                            & -1.995                                                   & 29.335                                                     & -0.957                                                 \\
{\color[HTML]{000000} CVaR(95\%)}                                                                               & -24.890                                                     & 49.413                                                 & -0.036                                                        & 0.697                                                             & -1.334                                                 & 0.519                                                            & -1.792                                                   & \textbf{27.676}                                            & -0.912                                                 \\
{\color[HTML]{000000} RL(Ret-based)}                                                                            & 33.573                                                      & 49.282                                                 & 0.053                                                         & 1.671                                                             & 1.021                                                  & 1.120                                                            & 1.523                                                    & 30.859                                                     & 1.502                                                  \\ 
{\color[HTML]{000000} R(Ours)}                                                                                  & \textbf{575.585}                                            & \textbf{52.840}                                        & \textbf{0.347}                                                & 4.181                                                             & \textbf{10.468}                                        & 3.031                                                            & \textbf{14.442}                                          & 65.519                                                     & \textbf{18.149}                                        \\ \hline
{\color[HTML]{000000} }                                                                                         & \multicolumn{9}{c}{\cellcolor[HTML]{FFFFFF}\textit{Port B}}                                                                                                                                                                                                                                                                                                                                                                                                                                                                                                           \\ \hline
{\color[HTML]{000000} MV}                                                                                       & -1.324                                                      & 49.078                                                 & -0.001                                                        & 0.461                                                             & -0.178                                                 & 0.456                                                            & -0.180                                                   & 11.408                                                     & -0.195                                                 \\
{\color[HTML]{000000} MAD}                                                                                      & -3.911                                                      & 48.896                                                 & -0.004                                                        & 0.458                                                             & -0.390                                                 & 0.465                                                            & -0.385                                                   & 11.739                                                     & -0.412                                                 \\
CVaR(95\%)                                                                                                      & -5.328                                                      & 49.192                                                 & -0.006                                                        & \textbf{0.447}                                                    & -0.522                                                 & 0.422                                                            & -0.553                                                   & 11.535                                                     & -0.547                                                 \\
RL(Ret-based)                                                                                                   & 20.384                                                      & 49.453                                                 & 0.026                                                         & 0.541                                                             & 1.368                                                  & \textbf{0.309}                                                   & 2.388                                                    & \textbf{9.961}                                             & 2.011                                                  \\
RL(Ours)                                                                                                        & \textbf{145.627}                                            & \textbf{51.730}                                        & \textbf{0.139}                                                & 1.815                                                             & \textbf{3.557}                                         & 1.187                                                            & \textbf{5.438}                                           & 18.370                                                     & \textbf{9.507}                                         \\ \hline
\end{tabular}
}
\end{center}
\end{table}

testing results reveal our model outperformed benchmark models, achieving total returns of 575.5\% and 145.6\% in Portfolios A and B, respectively, compared to non-converged return-based RL with a low positive return and SPPO models that only succeed in reducing market losses (equally weighted strategy). The non-convergence suggests that the return reward function is ineffective in guiding the agent to optimal solutions.

Despite facing higher risks due to increased portfolio value, our model demonstrated superior return-to-risk ratios, particularly in the Sortino and Kalmar ratios compared to the Sharpe ratio. This underscores its proficiency in managing downside risk. Additionally, the model excelled in the high-volatility period (Port A), illustrating its ability to capitalize on market fluctuations for higher profit while efficiently managing risk.

Table \ref{tab:4} also notes a nearly identical win rate (accuracy in predicting market direction) across all models. Thus, the marked difference in total returns can be attributed to the unique weight optimizations utilized by each model. RL models dynamically rebalance portfolio weights based on environmental insights and future event projections, prioritizing more profitable assets. In this context, our model exhibits greater accuracy than return-based RL due to convergence during the learning process. Conversely, SPPO benchmarks, which base weight rebalancing on historical data, often fail to adapt to market volatility, particularly in highly volatile markets like cryptocurrency.

\begin{figure}[H]
    \centering
    \includegraphics[scale=.76]{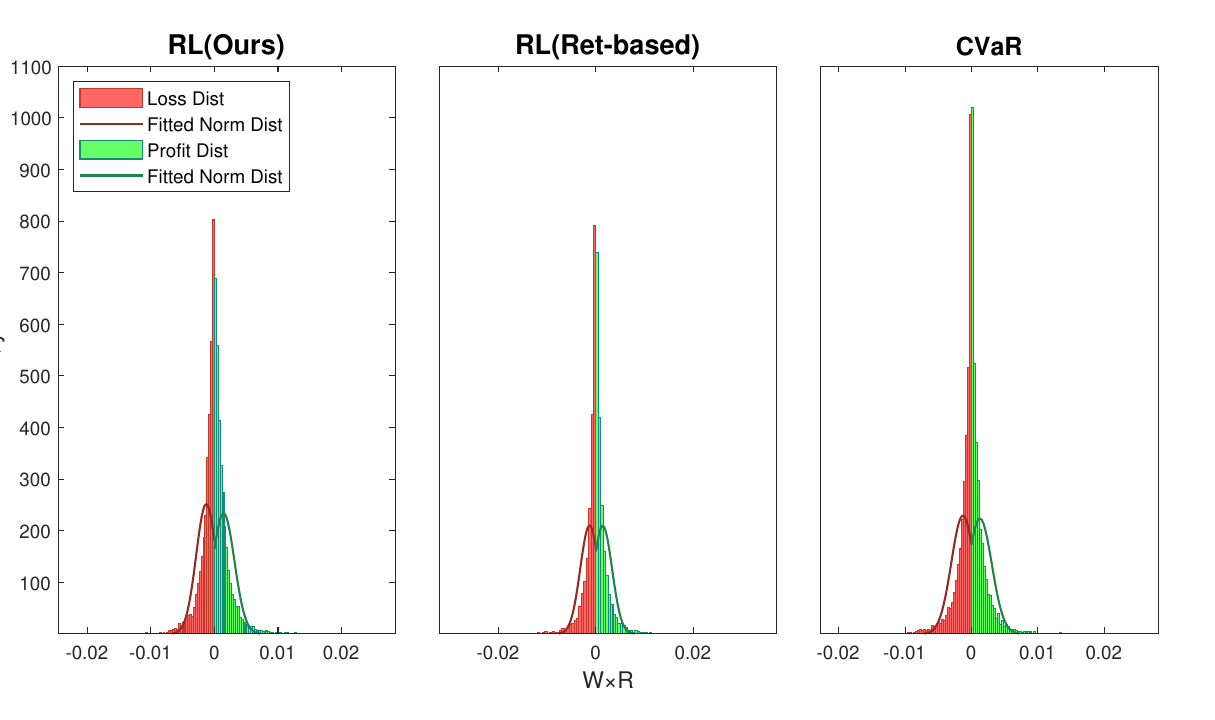}
    \caption{Distribution of $weight\times return$ and fitted normal distribution in Port A during testing.}
    \label{fig:7}
\end{figure}

\begin{figure}[H]
    \centering
    \includegraphics[scale=.76]{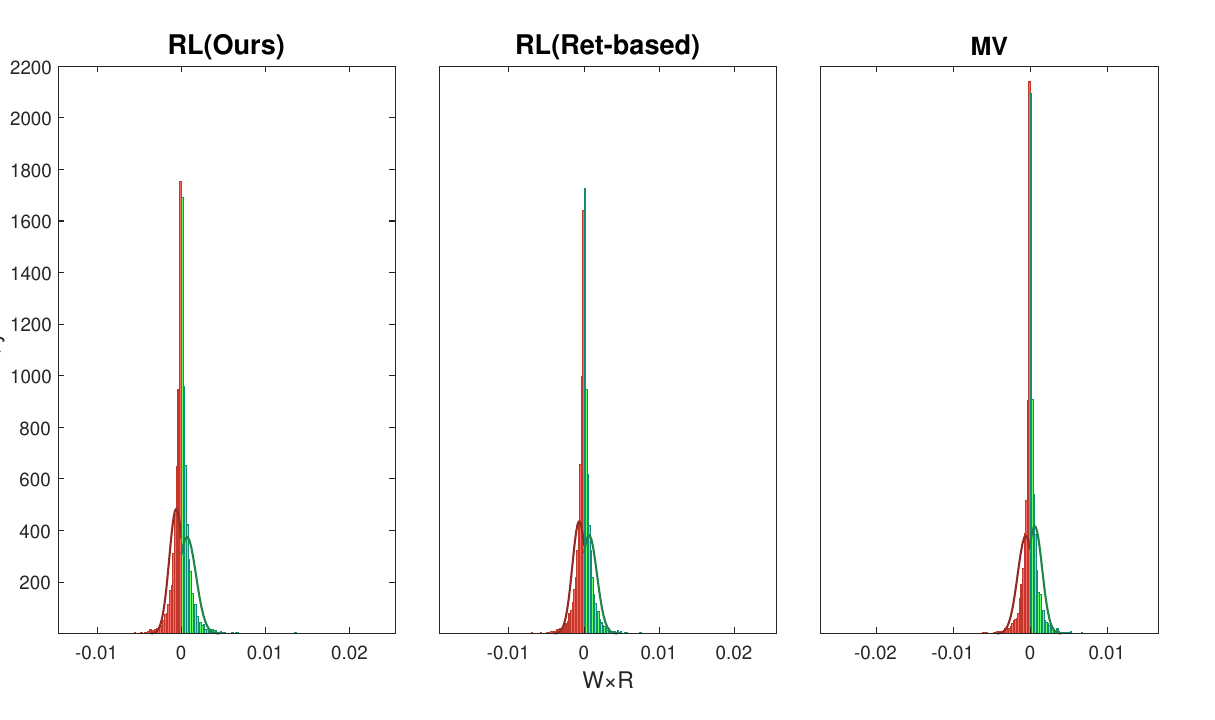}
    \caption{Distribution of $weight\times return$ and fitted normal distribution in Port B during testing.}
    \label{fig:8}
\end{figure}

Figures \ref{fig:7} and \ref{fig:8} complement this discussion by presenting the normal-curve-fitted distribution of the product of weights in the corresponding asset returns for our model alongside the top-performing benchmarks in Portfolios A and B during testing.

These distributions reveal a right-skewed profit distribution for our models, with kurtosis closer to zero for losses than profits. Conversely, the return-based RL, MV and CVaR models' distributions display kurtosis closer to zero for profits than for losses and left-skewed loss distribution, notably in the MV model.

\pagebreak

Therefore, despite having comparable win rates, benchmark models fail to achieve high profits and manage risk adequately.

\section{Conclusion} \label{Sec:conclusion}

In summary, this study introduces an RL-based portfolio management model tailored for high-risk settings, overcoming the limitations of traditional RL frameworks by leveraging two-sided transactions and lending. We developed a novel environmental formulation and a PnL-based reward function to improve the agent's convergence in training by enhancing its understanding of the additional risks presented by new features in the environment.

Testing results in the challenging cryptocurrency market indicate that our model surpasses benchmarks, including return-based RL, MV, MAD, and CVaR models, effectively managing a diversified 12-crypto asset portfolio across different market volatility. Additionally, our model demonstrated superior return-to-risk ratios, particularly in the Sortino and Kalmar ratios compared to the Sharpe ratio. This underscores its proficiency in managing downside risk. Furthermore, the model excelled during the high-volatility period, illustrating its ability to capitalize on market fluctuations for higher profit while efficiently managing risk.

This study highlights two main findings: First, our model adeptly incorporates two-sided transactions and lending in weight optimization to exploit all possible market opportunities, making it suitable for investors seeking profits under various market conditions. Second, by moving beyond the preset assumptions of SPPO models, our approach accommodates future asset price uncertainty, allowing for dynamic portfolio rebalancing and capturing the nonlinear effects of multiple asset risks to improve profitability.

Thus, the presented model offers a promising, state-of-the-art framework for investors aiming to achieve profit in all market conditions and secure higher return-to-risk ratios.

\pagebreak


\end{document}